\documentclass[preprint,review,12pt]{elsarticle}

\usepackage{graphics}
\usepackage{graphicx}

\usepackage{amssymb}
\usepackage{amsthm}
\usepackage{amsmath}
\usepackage{amsfonts}
\usepackage{amssymb}
\usepackage{amsthm}
\newtheorem{assumption}{Assumption}

\biboptions{sort&compress}

\journal{Physica A}

\begin{document}

\begin{frontmatter}


\title{Measuring the default risk of sovereign debt from the perspective of network}
\author[rvt]{Hongwei Chuang\corref{cor1}}
\ead{hongwei@stat.sinica.edu.tw}

\author[rvt,els]{Hwai-Chung Ho}

\cortext[cor1]{Corresponding author}

\address[rvt]{Institute of Statistical Science, Academia Sinica, Taiwan 11529}
\address[els]{Department of Finance, National Taiwan University, Taiwan 10617}

\begin{abstract}
Recently, there has been a growing interest in network research, especially in these fields of biology, computer science, and sociology. It is natural to address complex financial issues such as the European sovereign debt crisis from the perspective of network. In this article, we construct a network model according to the debt--credit relations instead of using the conventional methodology to measure the default risk. Based on the model, a risk index is examined using the quarterly report of consolidated foreign claims from the Bank for International Settlements (BIS) and debt/GDP ratios among these reporting countries. The empirical results show that this index can help the regulators and practitioners not only to determine the status of interconnectivity but also to point out the degree of the sovereign debt default risk. Our approach sheds new light on the investigation of quantifying the systemic risk. 
\end{abstract}


\end{frontmatter}


\section{Introduction}
The stability of the financial system and the potential of systemic risks to alter the functioning of this system have long been important for central banks and related research communities. Thus, guarding against systemic risk in the financial system is an emergent issue. However, specifically defining this type of risk and managing it is difficult. Therefore, the Federal Reserve Bank of New York and the National Research Council's Board on Mathematical Sciences and Their Applications held a conference to stimulate fresh ideas on systemic risk in May 2006 \cite{kambhu2007new}. The conference attracted more than 100 experts from 22 countries, representing banks, regulators, investment firms, US national laboratories, government agencies, and universities. This conference proposed the new directions for understanding systemic risk. A comprehensive survey of understanding systemic risk can be referred to \cite{bisias2012survey} and \cite{panizza2009economics}.  

Some external events, such as recessions, wars, civil unrest, environmental catastrophes or financial crisis, have the potential to depress the value of a banks' assets so severely that the system fails. In the wake of the global financial crisis that began in 2007, there is increasing recognition of the need to address risk at the systemic level, as distinct from focusing on individual banks \cite{billio2011econometric} \cite{haldane2009rethinking} \cite{haldane2011systemic}  \cite{jones2008preventing}. Sovereign debt defaults often make the financial system unstable, causing further systemic risk. A number of studies have been conducted on sovereign debt crises and the policy responses to these sovereign defaults \cite{roubini2004bailouts} following the sovereign debt crises in the 1980s. However, a little comparative empirical work has been done on the sovereign debt crises based on the macroeconomic and systemic perspectives. In this article, we provide an another view to analyze this kind of issue on sovereign debts from the network prospective.     

Currently, the increasingly complicated and globally interlinked financial markets are not immune to such systemic threats. Three questions immediately arise: (i) Does globalization make the world too interconnected? (ii) Is there a way to be immunized from the systemic threats? (iii) How can we find a proper index to measure systemic risk? In the following, we first focus on the role of growth in generating sovereign debt failure and instability. Second, we provide a systemic risk index to measure sovereign debt from the perspective of financial networks. Finally, we implement the empirical study through the quarterly data of consolidated foreign claims from the Bank for International Settlements (BIS) and debt/GDP ratios among the reporting countries. This index not only can provide the regulators and practitioners with the status of interconnectivity but also can point out the degree of the default risk. Moreover, it shields light on quantifying the systemic risk of the world.

\section{Connectivity of the sovereign debt}
To answer the question ``Does the globalization make the world too interconnected?,'' we use the quarterly report of consolidated foreign claims from BIS. The foreign claims by nationality of the reporting banks are the ultimate risk basis consisting of 20 countries from 2005 Q1 to 2011 Q1. The consolidated banking statistics reports banks' on-balance sheet financial claims on the rest of the world, thus providing a measure of the risk exposures of lenders' national banking systems. The quarterly data cover contractual lending by the head office and all its branches and subsidiaries on a worldwide consolidated basis.

We plot the debt--credit relations among these countries through FNA\footnote{http://www.fna.fi/}, as shown in Figures 1 and 2. FNA is an analytics platform that can help financial institutions and regulators better manage and understand financial data with network analysis and visualization. The nodes are represented as these 20 countries. The ties denote the debt--credit relations between any two of them. Thickness and thinness represent the debt--credit amount. 
Figure 1 shows the network structure of sovereign debts in 2005 Q1.
\begin{center}
$<$Insert Fig. 1 here$>$
\end{center}
Figure 2 shows that for 2009 Q4.      
\begin{center}
$<$Insert Fig. 2 here$>$
\end{center}
These countries became more inter-connective from 2005 Q1 to 2009 Q4. The connectivity of the world has indeed intensified during these past years' globalization.

\section{Network structure of sovereign debts}
The inability of previous approaches to reproduce statistical regularities observed empirically in network structures justifies our pursuit of a complex systems approach that may provide predictions for large-scale networks. Simple amplification mechanisms can dominate the network dynamics at large, despite the best intentions of the agents. Economic networks are subject to amplifications that may result from the redistribution of the load if one node fails. 

Here, we consider a network structure consisting of two parts: one part is the set of the nodes, $V$, in which $N$ countries belong,  and the other part is the sovereign debt--credit relationship among these countries, $E$. The network structure of these sovereign debts is formed and denoted by 
\begin{equation}
G \equiv (V, E).
\end{equation}

We further assume the following:
\begin{assumption}
For an arbitrary node $i \in V$, the default probability of node $i$ is defined as $p_i$.
 \end{assumption} 

\begin{assumption}
Let $q_{ij}$ be the probability that node $i$ defaults such that its linkage node $j$ defaults where $j \neq i$ and $q_{ii}=1, i=1, 2, ..., N$.
\end{assumption} 

\begin{assumption}
The loss of the node $i$ is $l_i$.
\end{assumption} 

According to these assumptions, the following properties can be immediately obtained:
\begin{enumerate}
\item The default transition matrix (DTM) of the network system is 
\[
\mbox{DTM}=
\begin{pmatrix}
1 & q_{12} & \ldots & q_{1N} \\
q_{21}& 1 & \ldots & q_{2N} \\
\vdots &  \vdots & \vdots & \vdots  \\
q_{N1} & \ldots & \ldots &1 \\
\end{pmatrix}
\ast Adj.(E), 
\]
where $Adj.(E)$ represents the adjacency matrix of $E$.
\item The network default probability (NDP) is
\[
\mbox{NDP}=
\mbox{DP
} \ast \mbox{DTM},
\]
where
\[
\mbox{DP}=
\begin{pmatrix}
p_1, &\ldots&, p_N
\end{pmatrix}.
\] 
\item The network expected loss (NEL) is
\[
\mbox{NEL}=
\mbox{NDP} \ast \mbox{L},
\]
where
\[
\mbox{L}=
\begin{pmatrix}
l_1, &\ldots&, l_N
\end{pmatrix}'.
\] 
\end{enumerate}

\section{Sovereign debts and systemic risk}
Although systemic risk is a difficult concept to define precisely, a better understanding of systemic risk is given by \cite{schweitzer2009economic}: ``If a single node fails, it may force other nodes to fail as well, which may eventually lead to failure cascades and the breakdown of the system, denoted as systemic risk.'' 

According to the Property Casualty Insurers Association of America (PCIAA), there are two key assessments for measuring systemic risk: the ``too big to fail'' (TBTF) and the ``too interconnected to fail'' (TICTF) tests\footnote{http://www.pciaa.net/}. The TBTF test is the traditional analysis for assessing the risk of required government intervention. TBTF can be measured in terms of an institution's size relative to the national and international marketplaces, market share concentration, and competitive barriers to entry or how easily a product can be substituted. The TICTF test is a measure of the likelihood and amount of medium-term net negative effect on the larger economy of an institution's failure to conduct its ongoing business. The effect is measured beyond the institution's products and activities to include the economic multiplier of all other commercial activities dependent specifically on that institution. The effect is also dependent on how correlated an institution's business is with other systemic risks.
 
Based on the two principles of PCIAA, finding such few economic variables to describe $DP$, $DTM$ and $L$ are difficult. We adopt two economic variables which can be observed to proxy them and also determine the systemic risk of these sovereign debts. We define systemic risk index (SRI) as
\begin{equation} 
\mbox{SRI}\equiv \sum_{i=1}^N d_i \cdotp V(k_i),
\end{equation}
where $d_i$ and $V(k_i)$ represent the debt/GDP ratio and the functions of the topological importance of node $i$, respectively. To describe the topological importance of node $i$, the network centrality is proper. There are two kinds of centrality measurements commonly be used in network theory: local measure and non-local measure. The local measure indicates the degree centrality or closeness centrality while the non-local measure denotes the betweenness centrality or eigenvector centrality. Here, we use betweenness centrality to measure  the topological importance of node $i$. The betweenness centrality of a node $i$, $g(i)$, is given by the following expression:  
\begin{equation} 
g(i)=\sum_{s\neq v\neq t}\frac{\sigma_{st}(i)}{\sigma_{st}},
\end{equation}
where $\sigma_{st}$ is the total number of shortest paths from node $s$ to node $t$, and $\sigma_{st}(i)$ is the number of the paths that pass through $i$. We plot the time series of SRI from 2005 Q1 to 2011 Q1 as shown in Figure 3.
\begin{center}
$<$Insert Fig. 3 here$>$
\end{center}

Figure 3 indicates that there is a peak around 2010 Q2. Beginning late 2009, fears of a sovereign debt crisis developed among investors as a result of the rising government debt levels around the world along with a wave of downgrading of government debts in some European states. Concerns intensified in early 2010 and thereafter, causing Europe's finance ministers on May 9, 2010 to approve a rescue package worth 750 billion euros aimed at ensuring financial stability across Europe by creating the European Financial Stability Facility. 
 
\section{Conclusions}
\cite{ibragimov2011diversification} argues the argument that the recent financial crisis has significant externalities and systemic risks arising from the interconnectedness of financial intermediary risk portfolios. The negative externality arises because intermediaries' actions to diversify that are optimal for individual intermediary may prove to be suboptimal for the society. This externality depends critically on the distributional properties of the risks. The optimal social outcome involves less risk--sharing, but also a lower probability for the massive collapse of intermediaries. Furthermore, \cite{podobnik2010time} and \cite{wang2011quantifying} study the time-lag cross-correlations in multiple time series by using time-lag random matrix theory. The increase in the level of globalization is related with the increase in cross-correlations between different financial indices. The magnitude of the cross-correlations constitute ``bad news'' for the international investment managers who may believe the risk is reduced by diversifying across countries. 

In this paper, we propose a new index to measure the systemic risk in financial systems. We use the quarterly report of consolidated foreign claims from the
BIS and debt/GDP ratios among these reporting countries. Our result indicate that the index can really help to quantify the level of systemic risk. The index could further be considered in a wide range of global market or other network-typed financial systems. 

\section*{Acknowledgements}
This research is supported by the National Science Council  (NSC-100-2118-M-001-007-MY2) in Taiwan. We are grateful to the seminar participants at Institute of Sociology, Academia Sinica for their helpful comments and suggestions. We also thank Kimmo Soramaki and the team of FNA for their software support. We thank the anonymous referees for their valuable comments. All errors are our own.





\bibliographystyle{model1c-num-names}
\bibliography{PHYSA}






\newpage
\begin{figure}
\begin{center}
   \includegraphics[width=0.8\textwidth]{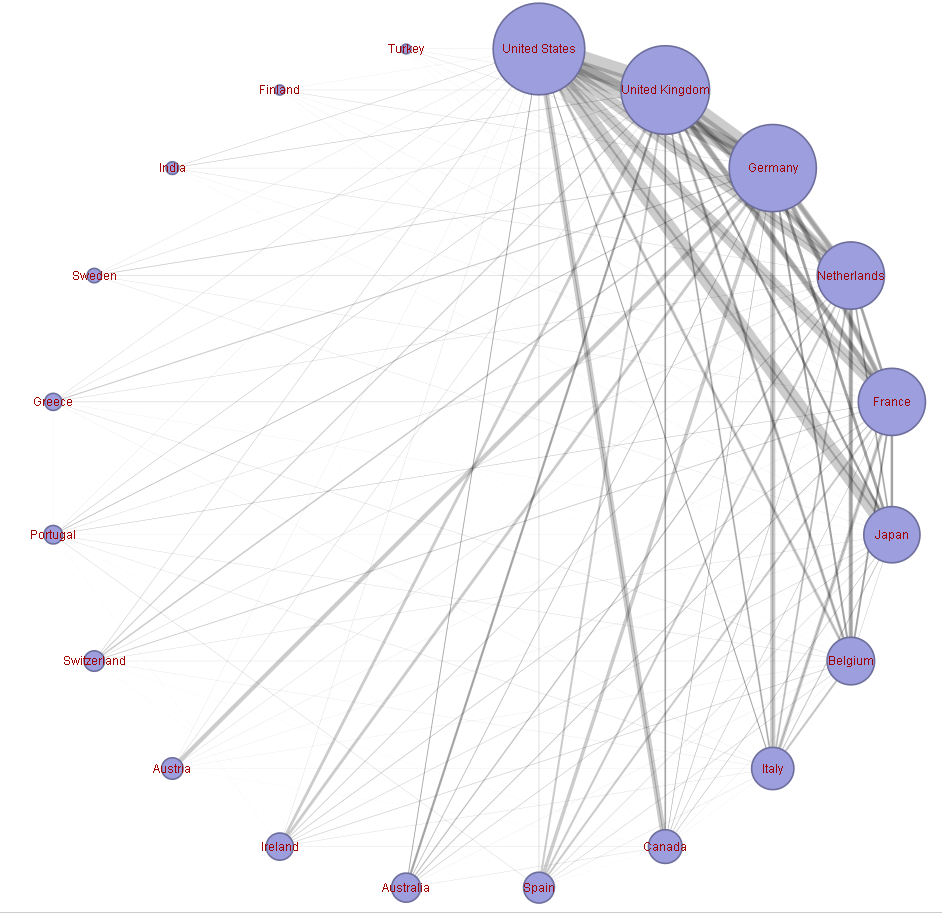}
   \caption{Network Structure of Sovereign Debt in 2009 Q4}
\end{center}
\end{figure}

\begin{figure}
\begin{center}
   \includegraphics[width=0.8\textwidth]{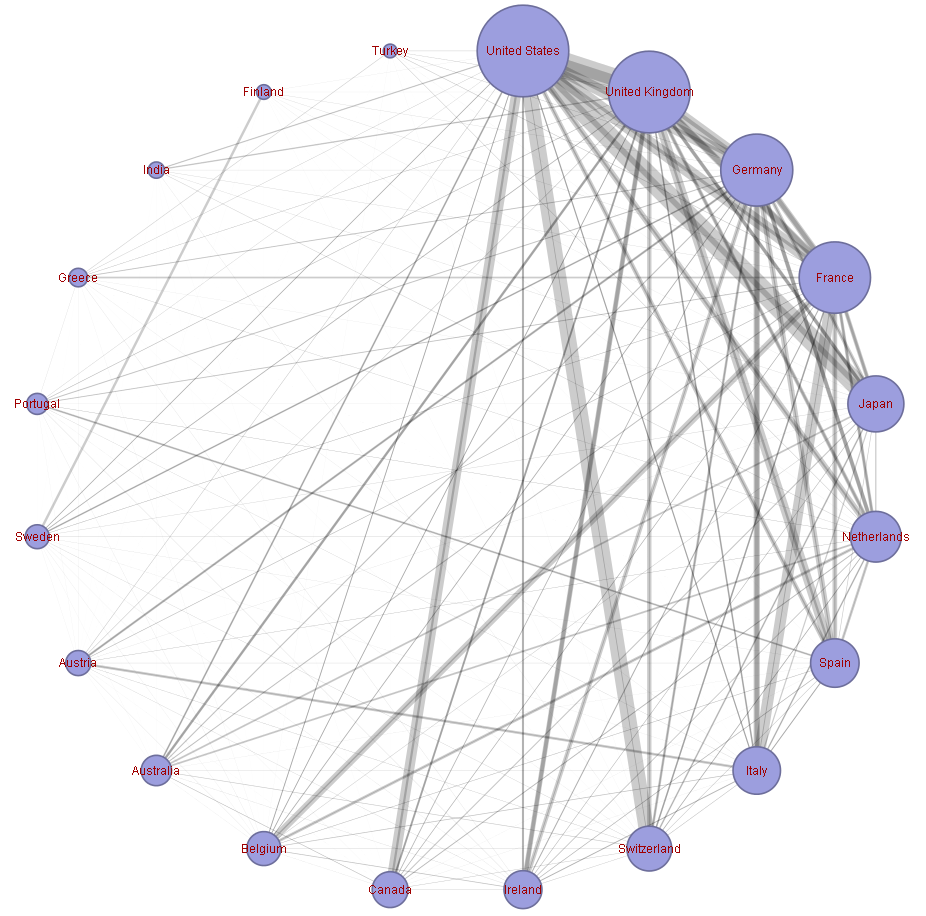}
   \caption{Network Structure of Sovereign Debt in 2009 Q4}
\end{center}
\end{figure}

\begin{figure}
\begin{center}
   \includegraphics[width=0.95\textwidth]{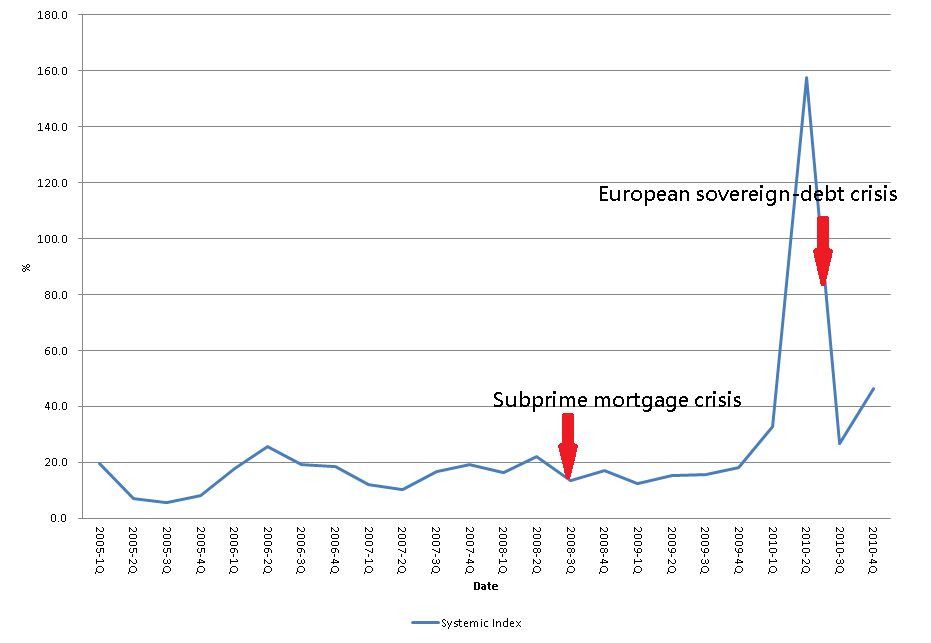}
   \caption{Systemic Risk Index}
\end{center}
\end{figure}

\end{document}